\documentstyle[12pt,psfig]{article}
\def\beq{\begin{equation}}
\def\eeq{\end{equation}}
\def\bea{\begin{eqnarray}}
\def\eea{\end{eqnarray}}
\def\bq{\begin{quote}}
\def\eq{\end{quote}}

\def\gappeq{\mathrel{\rlap
{\raise.5ex\hbox{$>$}}
{\lower.5ex\hbox{$\sim$}}}}

\def\lappeq{\mathrel{\rlap{\raise.5ex\hbox{$<$}}
{\lower.5ex\hbox{$\sim$}}}}

\begin{document}
\pagestyle{empty}
\begin{flushright}
{CERN-TH/97-343\\
IFT-97/19\\
hep-ph/9712234}
\end{flushright}
\vspace*{5mm}
\begin{center}
{\bf
THE FINE-TUNING PRICE OF LEP} \\
\vspace*{1cm} 
 Piotr H. Chankowski$^{a)}$,
John Ellis$^{b)}$ and Stefan Pokorski$^{a,b)}$
\\
\vspace{0.3cm}
\vspace*{1.5cm}  
{\bf ABSTRACT} \\
\end{center}
\vspace*{5mm}
\noindent
We quantify the amount of fine tuning of input parameters of the Minimal
Supersymmetric Extension of the Standard Model (MSSM) that is needed to 
respect the lower limits on sparticle and Higgs masses imposed by precision 
electroweak measurements at LEP, measurements of $b \rightarrow X_s \gamma$, 
and searches at LEP 2.  If universal input scalar masses are assumed in a 
gravity-mediated scenario, a factor of $\gappeq180$ is required at 
$\tan\beta\sim1.65$, decreasing to $\sim20$ at $\tan\beta \sim 10$. The
amount of fine tuning is not greatly reduced if non-universal input scalar 
Higgs masses are allowed, but may be significantly reduced if some 
theoretical relations between MSSM parameters are assumed.

\vspace*{5cm} 
\noindent

\rule[.1in]{16.5cm}{.002in}

\noindent
$^{a)}$ Institute of Theoretical Physics, Warsaw University.\\
$^{b)}$ Theory Division, CERN, Geneva, Switzerland.
\vspace*{0.5cm}

\begin{flushleft} CERN-TH/97-343
\\
IFT-97/19 \\
November 1997
\end{flushleft}
\vfill\eject
%\pagestyle{empty}
%\clearpage\mbox{}\clearpage

\setcounter{page}{1}
\pagestyle{plain}

The measured magnitudes of the gauge coupling strengths are consistent with the
presence of  light sparticles within a supersymmetric GUT model~\cite{SGUT}, 
- for a recent updated analysis see \cite{CHPLPO} -
and precision electroweak data are also consistent with a relatively light 
Higgs boson~\cite{EWHIGGS,LEPEWWG} as predicted in the MSSM, but other 
{\it ad hoc} 
interpretations of these pieces of circumstantial evidence for low-energy 
supersymmetry are also possible. The primary phenomenological motivation for 
observable supersymmetric particles with masses $\lappeq$~1~TeV is that they 
render $M_W \ll M_{Pl}$ natural and thereby alleviate the fine tuning of input
parameters required to keep $M_W$ small. Some time ago, it was proposed 
\cite{FINETUNE,BAGI} that the amount of fine tuning be measured by the
logarithmic
sensitivity $\Delta$ of $M_Z$ to input model parameters $a$: $\Delta\equiv
\left\vert(a / M_Z^2 (\partial M_Z^2 / \partial a)\right\vert$,
and
the requirement that $\Delta$ be not too large was used to motivate
numerically the lightness of sparticles~\cite{FINETUNE,BAGI,ANCA,DIGI}.

This argument offered hope that some sparticles might be detected at LEP~2.  
At the time of writing, no such sparticles have been seen, nor have any Higgs
bosons~\cite{JANOT,LEPC183}, and precision electroweak measurements 
\cite{WARD} and observations of $b\rightarrow X_s\gamma$ decay \cite{CLEO} 
are consistent with the Standard Model. This depressing lack of evidence for 
supersymmetry is in {\it prima facie} disagreement with some of the previous 
optimistic suggestions~\cite{BAGI,DIGI} motivated by the absence of fine 
tuning. How much should one worry about this apparent disappointment? The 
answer is necessarily subjective, since the fine-tuning argument is not a 
rigorous mathematical statement, but rather an intuitive physical preference. 
However, it is possible to make an objective contribution to the debate by 
quantifying the amount of fine tuning that is required by the data. The
reader 
may then reach her/his own judgement how seriously to take the continued 
absence of supersymmetry.

This paper describes a first attempt to formulate the fine-tuning problem in
this way. Our theoretical framework is that of supergravity with
gravity-mediated supersymmetry breaking and universal gaugino masses
$M_{1/2}$ and trilinear (bilinear) supersymmetry-breaking parameters
$A_0 \; (B_0)$ at the input
supergravity scale~\footnote{Similar 
considerations may also be applied to gauge-mediated models, but lie beyond 
the scope of this paper.}. 
We shall for the most part assume universality also for 
the input scalar masses $m_0$, but shall also discuss the implications of 
relaxing this assumption for the Higgs scalar masses.
The data we take into account include the latest 
set of precision electroweak data reported at the Jerusalem conference
\cite{WARD}, which are dominated by those from LEP~1, the latest measurement 
of $B(b\rightarrow X_s\gamma)$  by the CLEO collaboration \cite{CLEO}, and 
the lower limits on sparticle and Higgs boson masses from LEP~2. For the 
latter, we again base ourselves on the data reported in Jerusalem \cite{JANOT},
but also comment on the impact of more recent limits from LEP running at
183~GeV \cite{LEPC183}. To set our results in context, we also remark on
the inflation in the price of fine tuning since the initial LEP runs in
1990, and mention the potential implications of non-observation of
supersymmetry when LEP~2 running is completed, and if no sparticles
appear during Run~II of the Tevatron. 

At the present time, we find that a fine-tuning price $\Delta\gappeq180$ must
be paid if $\tan\beta$ is close to its infra-red fixed-point value and 
universal boundary conditions are chosen for the input scalar masses $m_0$. 
This price is reduced to $\Delta\simeq60$ for $\tan\beta=2.5$, and 
$\Delta\simeq20$ for $\tan\beta=10$. The fine-tuning price is not
decreased significantly if one allows the input scalar Higgs masses to be 
non-universal, because there are additional parameters whose fine tuning
must be taken into account in evaluating $\Delta$. In the absence  of an 
objective criterion for interpreting $\Delta$, we observe that
$\Delta\sim3$ 
was possible before LEP started setting limits on supersymmetry, and that if 
the remaining stages of LEP~2 do not find the lightest supersymmetric
Higgs
boson with a mass below 95~MeV, there will be lower bound $\Delta\gappeq1000$ 
for $\tan\beta\simeq1.65$ and $\Delta\gappeq130$ for $\tan\beta = 2.5$,
though the 
impact will be less severe for larger values of $\tan\beta$. For higher
values of $\tan\beta(\sim10)$, the minimal 
amount of fine tuning is for $M_h\approx 105$ GeV. 
The non-observation of gluinos and squarks at the FNAL Tevatron collider 
during Run~II would not increase the fine-tuning price much further.
Non-universal boundary conditions for the Higgs scalar masses do not
reduce greatly the fine-tuning price, but it could be reduced
significantly if there was some 
theoretical relation between the input MSSM parameters.

Before discussing our analysis in more detail, we first specify more precisely
the fine-tuning criterion we use.  Following \cite{FINETUNE,BAGI,DIGI}, we 
consider the logarithmic sensitivities of $M_Z$ with respect to variations in 
input parameters $a_i$:\
\beq
\Delta_i=\left\vert{a_i\over M_Z^2}~ 
         {\partial M_Z^2\over\partial a_i}\right\vert
\label{eqn:e1}
\eeq
and then define
\beq
\Delta = {\rm max}_i \Delta_i
\label{eqn:e2}
\eeq
In the specific case of the MSSM with universal gaugino and scalar masses
$(M_{1/2},m_0)$  at the input supergravity scale and a  universal trilinear
(bilinear) supersymmetry-breaking parameter $A_0 \; (B_0)$, we consider
the following input
parameters $a_i$:
\beq
(M_{1/2}, m_0, \mu_0, A_0, B_0)
\label{eqn:e3}
\eeq
and we use the tree level formula for the scalar Higgs potential, with
parameters renormalized at the electroweak scale~\footnote{The full
one-loop 
corrections to the scalar potential relax the degree of fine tuning by 
20-30\% ~\cite{OLPO}. We ignore this effect here, since it is inessential
for our 
conclusions.}. The $\Delta_i$ are calculated as in ref. \cite{DIGI}, with
the 
dependence of $\tan\beta$ on the input parameters taken into account, from 
the master formula
\begin{eqnarray}
\Delta_i = \left\vert{2a_i\over(\tan^2\beta-1)M^2_Z}
\left\{{\partial m^2_1\over\partial a_i} - 
\tan^2\beta{\partial m^2_2\over\partial a_i}
-{\tan\beta\over\cos2\beta}\times\right.\right.\nonumber\\
\left.\left.
\left(1+{M^2_Z\over m^2_1+m^2_2}\right)
\left[2{\partial m^2_3\over\partial a_i}
-\sin2\beta\left({\partial m^2_1\over\partial a_i}
+ {\partial m^2_2\over\partial a_i}\right)\right]\right\}\right\vert
\label{eqn:e4}
\end{eqnarray}
where the $m^2_i$ are the mass parameters of the Higgs potential of the
MSSM.

We now review in more detail the data set used in our analysis. As already 
mentioned, we use
the precision electroweak data set reported at the Jerusalem conference
\cite{WARD,LEPEWWG}. As is well known, the data set are fitted well by the 
Standard Model with a value of the Higgs mass compatible with MSSM predictions,
and measurements of $Z^0\rightarrow\bar b b$, $\bar c c$ decays no longer give 
any hint of new physics beyond the Standard Model. We constrain MSSM parameters
by requiring that $\Delta\chi^2<4$ in a global MSSM fit \cite{MSSMFITS}. The 
main effect of this constraint is a lower bound on the left-handed stop, 
$M_{\tilde t_L}\gappeq300-400$ GeV \cite{CHPO,KANE2}. We also take into
account 
the direct LEP~2 lower limits on the masses of sparticles and Higgs bosons 
that were also reported at Jerusalem. Qualitatively, these impose 
$M_{1/2}\gappeq100$~GeV but still alow $m_0\rightarrow0$ in the absence of 
other constraints. As we shall see, an important r\^ole can played by searches 
for MSSM Higgs bosons. However, the preliminary results from data taken around
183~GeV in centre-of-mass energy, although representing a significant advance 
on the Jerusalem data by imposing $M_h\gappeq 75$~GeV, are still insufficient 
to increase the fine tuning price beyond that already required by the rest of 
the constraints. For that, one must wait for further upgrades of the LEP~2 
energy.

The final accelerator contraint we use is the measured value of 
$1\times10^{-4}<B(b \rightarrow X_s\gamma) <4.2\times10^{-4}$ at 95\% C.L.
\cite{CLEO}. The interpretation of this measurement in the MSSM is still 
subject to some uncertainty, because not all the ${\cal O}(\alpha_s)$ 
corrections have yet been calculated. Resumming large QCD logaritms up to 
next-to-leading order (NLO) accuracy has been recently accomplished in the 
SM \cite{ADYA}. All these calculations are identical in the SM and the MSSM 
except for that the initial numerical values of the Wilson coefficients at 
the scale $\mu\approx M_W$ are different. In our analysis we have used for 
them only the leading order results available in the MSSM. The uncertainty
due to order $\alpha_s/\pi$ corrections to them has been, however,
included as in ref. \cite{MIPORO,KANE2}. Those references also contain 
extensive discussion of the role played by the $b\rightarrow s\gamma$ 
measurement in constraining the parameter space of the MSSM. 

An important r\^ole may also be played by non-accelerator constraints, in
particular the relic cosmological density of neutralinos $\chi$, if these are
assumed to be the lightest supersymmetric particles, and if $R$ parity is
absolutely conserved.  Both of these assumptions may be disputed, and a
complete investigation of astrophysical and cosmological constraints is beyond
the scope of this analysis.  We limit ourselves to a qualitative discussion
based on the requirement that $0.1\leq\Omega_\chi h^2\leq0.3$, where
$\Omega_\chi$ is the density of neutralinos relative to the critical density,
and $h$ is the present Hubble expansion rate in units of
100~kms$^{-1}$Mpc$^{-1}$. Previous discussions \cite{ELFAOLSCH} have indicated
that this requirement can be satisfied for some parameter choices
in the ranges $0.2\lappeq m_0/M_{1/2}\lappeq1$
and $M_{1/2}\lappeq450$~GeV. We comment later on the potential impact of these
constraints.

We illustrate our discussion of fine tuning by discussing three specific
choices of $\tan\beta$: 1.65, which is favoured by an infra-red fixed-point
analysis and on the verge of being excluded by a more detailed analysis of the
compatibility between accelerator and astrophysical constraints, an
intermediate choice $\tan\beta=2.5$, and a higher value $\tan\beta=10$. The
discussion of larger values of $\tan\beta$ requires a 
more complete treatment of the
renormalization-group equations below the supergravity scale, which is beyond
the scope of this paper.

\begin{figure}
\psfig{figure=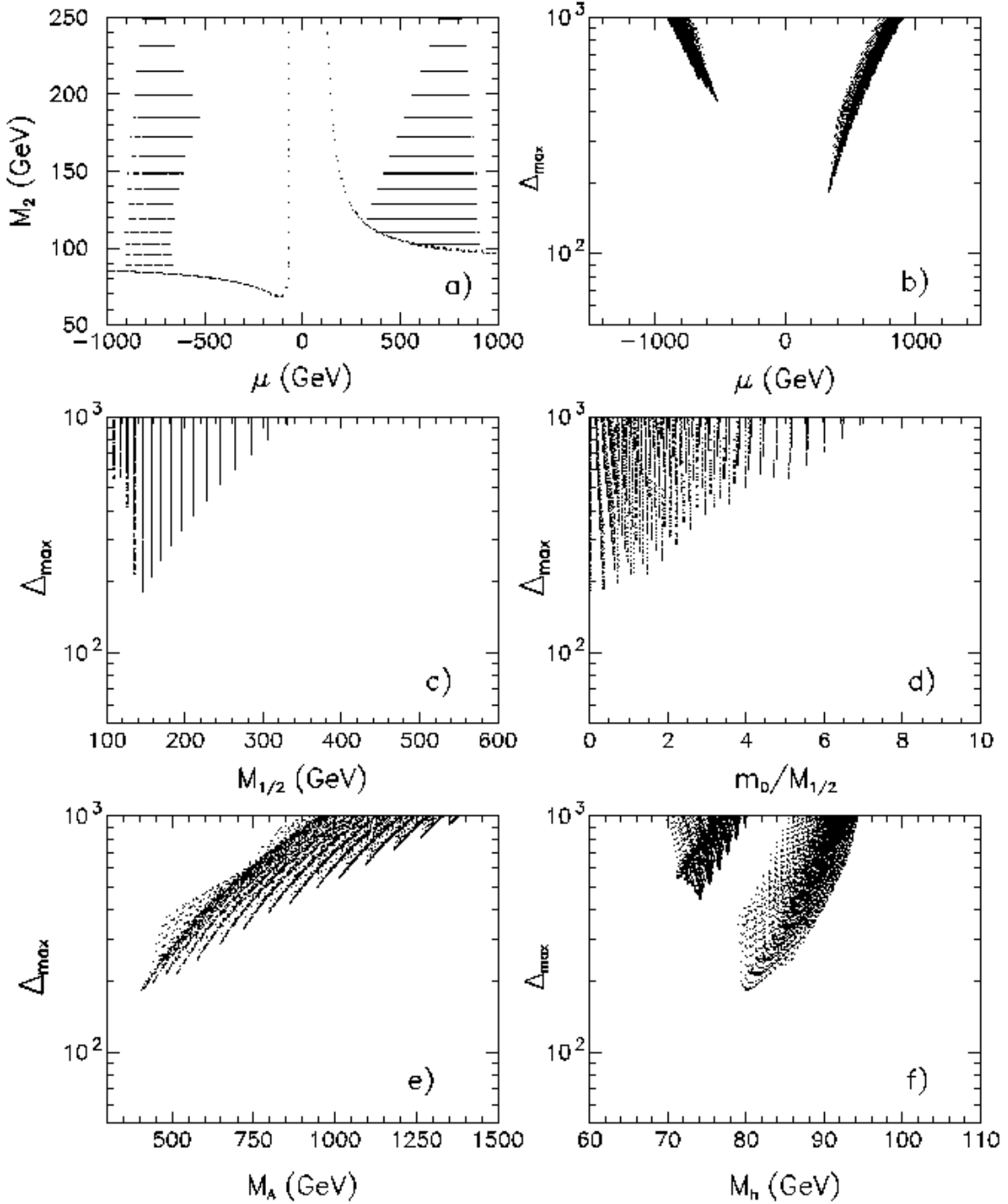,height=19.0cm}
\caption{The price of fine tuning for tan$\beta = 1.65$, assuming universal
input scalar masses at the supergravity scale. Panel (a) displays the
regions of the $(\mu, M_2)$ plane that are allowed by LEP~2, and the
restricted regions permitted when the other constraints discussed in
the text are implemented. The other panels display the ranges of the
fine-tuning parameter $\Delta$ obtained as functions of 
(b) the Higgs mixing parameter $\mu$, (c) the input
gaugino mass parameter $M_{1/2}$, (d) the ratio of the universal scalar
mass $m_0$ to $M_{1/2}$, (e) the CP-odd neutral Higgs mass $M_A$,
and (f) the lightest neutral Higgs mass $M_h$.}
\label{fig:f1}
\end{figure}

The case $\tan\beta=1.65$ with universal input scalar masses is displayed in
Fig.~\ref{fig:f1}. Panel (a) shows the $(\mu,M_2)$ plane, including the 
boundaries of the regions excluded by direct LEP searches for charginos and 
neutralinos now and at LEP~1. We see that the combination of the requirement 
of the proper electroweak breaking (which is possible only for 
$\mu>M_{1/2},m_0$), precision electroweak data and the $b\rightarrow
X_s\gamma$
constraint disallow regions of low $\mu$ and $M_2$ that were not excluded by 
the direct searches, particularly for $\mu<0$.  
Panel (b) exhibits a strong correlation between $\Delta$
and $\vert\mu\vert$, that $\Delta$ increases as $\vert\mu\vert$ 
increases, and that more fine tuning is required for
negative values of $\mu$. 
Panel (c) displays possible
values of $\Delta$ versus values of $M_{1/2}$.  We see that the
minimal values of $\Delta$ are for $M_{1/2}\approx 140$ GeV, and 
that they increase 
rapidly for smaller or larger $M_{1/2}$. The increase for increasing $M_{1/2}$ 
has an obvious reason, whereas that at low values of $M_{1/2}$ is due to the 
constraints discussed above, which in that case require a larger value
of $m_0$. Panel (d) 
displays values of $\Delta$ versus the ratio $m_0/M_{1/2}$, where we see little
dependence for $m_0/M_{1/2}<2$, whilst the fine-tuning price increases for
larger 
values of this ratio. Panel (e) shows a correlation of $\Delta$ with
the CP-odd neutral Higgs mass $M_A$: lower values of $M_A$ are
disallowed by the $b \rightarrow X_s \gamma$ constraint.
Finally, panel (f) shows the variation of $\Delta$ with the 
mass of the lightest MSSM Higgs boson $M_h$. The two populated regions
correspond to the different signs of $\mu$: since $A_t$ is essentially
determined by $M_{1/2}$ in the neighbourhood of the fixed point, these
different signs correspond to different amounts of $\tilde t$ mixing,
and hence different ranges of $M_h$. We see that $\Delta$ increases with 
$M_h$, as might be expected from the sensitivity of $M_h$ to $m_0$ and 
$M_{1/2}$ via radiative corrections, and the dependences of $\Delta$ on 
$M_{1/2}$ and $m_0$ seen in panels (c) and (d).  As the LEP~2 energy
increases,
and correspondingly the experimental sensitivity to $M_h$, continued
non-observation of the lightest MSSM Higgs boson would increase significantly 
the fine-tuning price imposed by LEP.

\begin{figure}
\psfig{figure=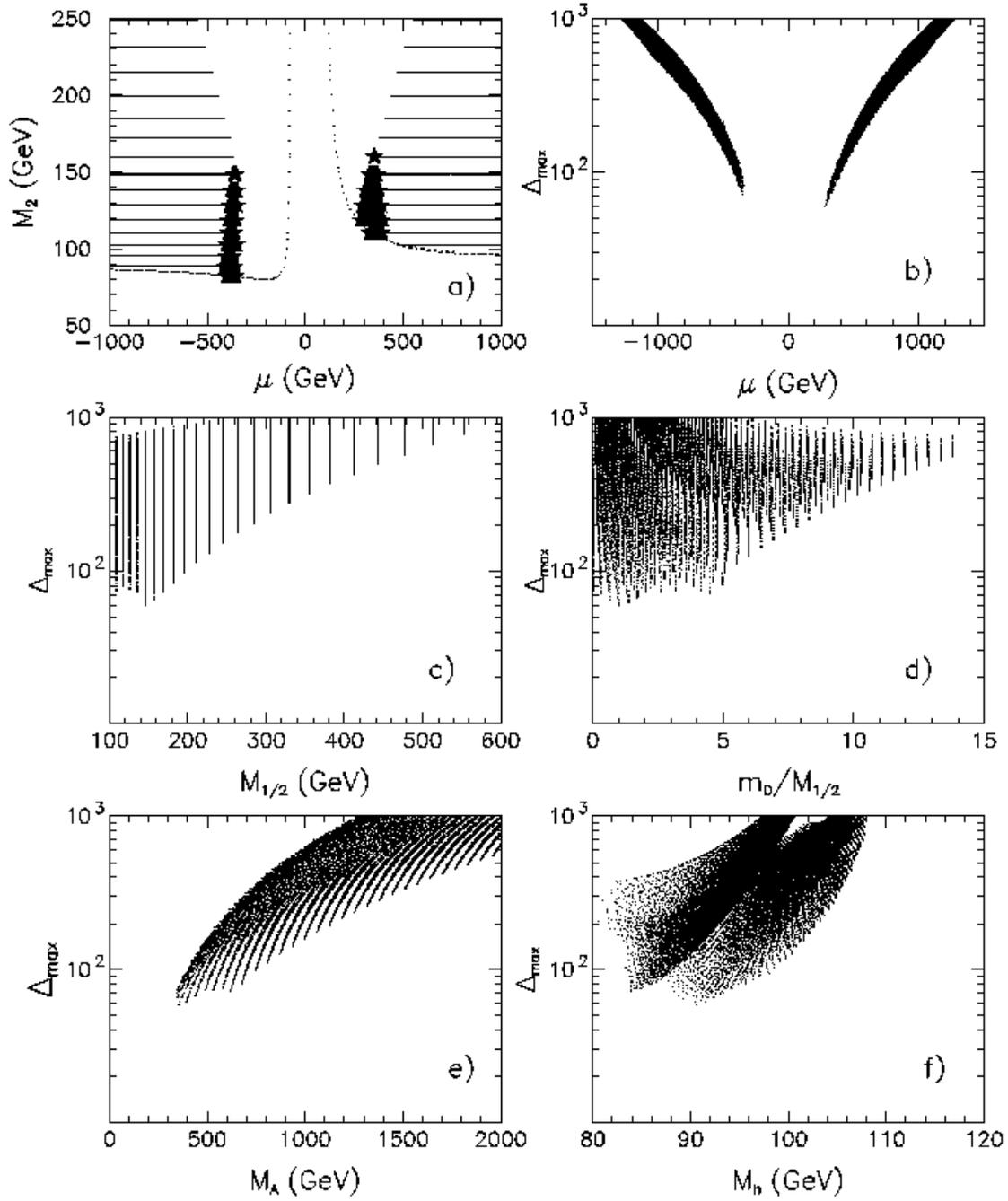,height=19.0cm}
\caption{As for Fig.~1, but for the value $\tan\beta = 2.5$.
Black stars in panel (a) correspond to points with $\Delta<100$.}
\label{fig:f2}
\end{figure}

Figure~\ref{fig:f2} displays corresponding panels for the choice 
$\tan\beta=2.5$. Panel (a) shows that the non-LEP constraints exclude
a smaller region of the $\mu, M_{1/2}$ plane for negative $\mu$ than
was the case for smaller tan$\beta$. This is reflected in a reduction
in the minimum fine-tuning price
to $\Delta\simeq60$, as we see in the other panels. We see that this 
is attained when $M_{1/2} \sim 100-140$~GeV [panel (c)]
and also $m_0 \sim 300-400$~GeV corresponding to a
relatively large value of $m_0/M_{1/2}\sim3-4$ [panel (d)]. Finally, we note 
in panel (f) that the minimum value of $\Delta$ is attained when
$M_h\sim85$~GeV, beyond the current reach of LEP but accessible to future
LEP~2 upgrades.
Beyond this value of $M_h$,
the fine-tuning price increases significantly, though it is always less than 
for $\tan\beta=1.65$, reflecting less need for large values of $m_0$ and 
$M_{1/2}$ to yield the same value of $M_h$ via radiative corrections.

\begin{figure}
\psfig{figure=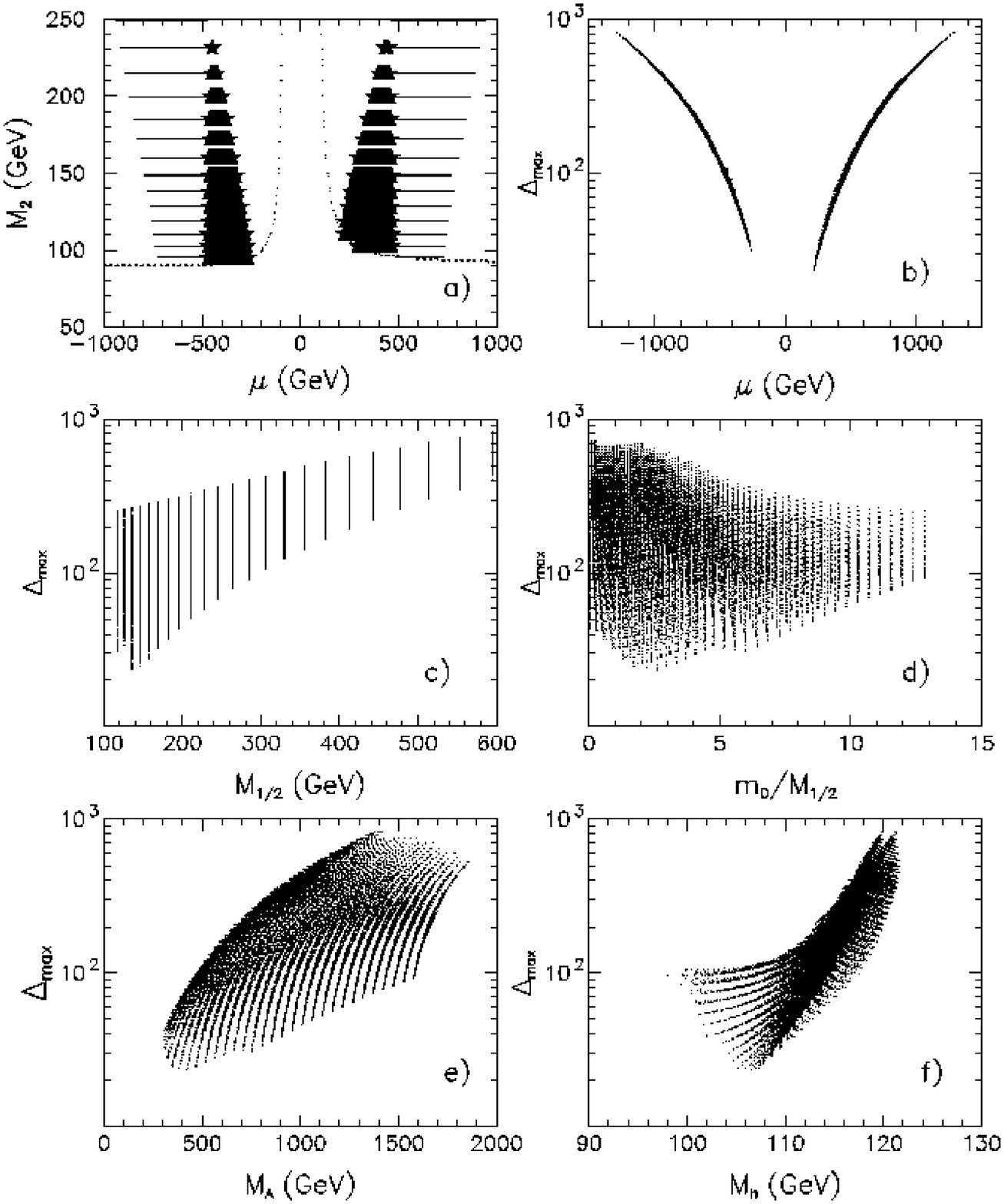,height=19.0cm}
\caption{As for Figs.~1 and 2, but for the value $\tan\beta = 10$.
Black stars in panel (a) again correspond to points with $\Delta<100$.}
\label{fig:f3}
\end{figure}

The corresponding analysis for $\tan\beta=10$ is displayed in 
Fig.~\ref{fig:f3}. We see in panel (b) that the correlation
between $\Delta$ and $\mu$ has now become very tight, and
note in panel (c) a familiar tendency for $\Delta$ to 
increase with $M_{1/2}$, once a minimum around 140~GeV has been passed. The 
minimum in panel (d) is for $m_0/M_{1/2}\sim2$ to 5 and it is somewhat more 
pronounced than for smaller $\tan\beta$.  Panel (b) shows the same correlation
between $\Delta$ and $\mu$ as for other values of $\tan\beta$. Finally,
we see in panel (f) that $\Delta$ is minimized when $M_h\sim105$ to 110~GeV,
which is probably beyond the reach of LEP~2.

\begin{figure}
\psfig{figure=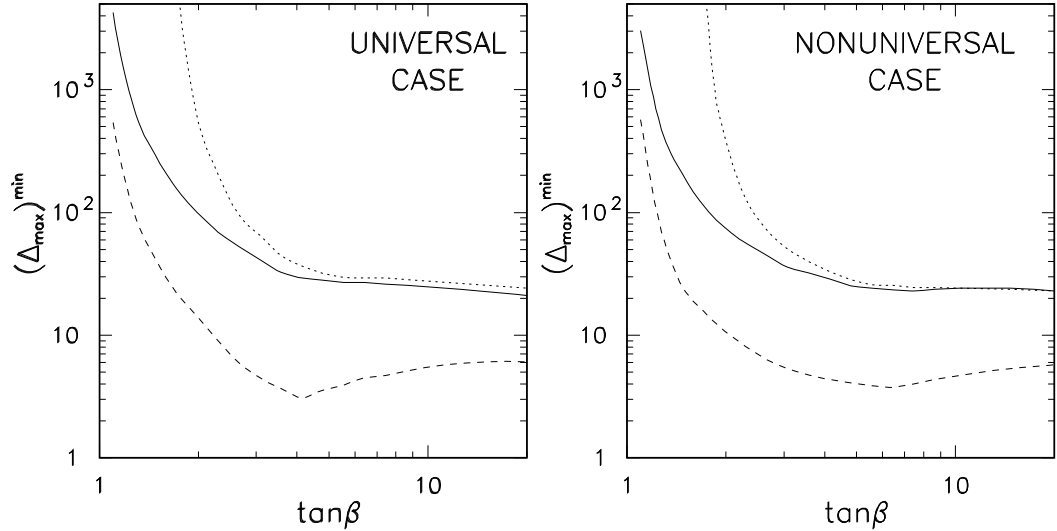,height=8.0cm}
\caption{Compilation of the minimal values of the fine-tuning parameter 
$\Delta$ as a function of tan$\beta$. The left panel is for the case of 
universal scalar masses, with the current constraints from LEP, 
$b\rightarrow X_s\gamma$, etc., shown as a solid line. The dashed line is for 
the constraints that were available after the initial runs of LEP~1, and the 
dotted line indicates what might be the situation if no evidence for 
sparticles or MSSM Higgs bosons is found with future upgrades of LEP~2. The 
right panel shows the corresponding lower limits on $\Delta$ for the case 
of non-universal Higgs masses, using the same conventions for the lines.}
\label{fig:f4}
\end{figure}

Figure~\ref{fig:f4}  assembles our information on the minimum value of 
$\Delta$ as a function of $\tan\beta$.  The current lower limit, assuming 
universal input scalar masses and the current data set reviewed earlier, is 
shown in the left panel
as a solid line. The fine-tuning price is not strongly dependent on
$\tan\beta$, except for $\tan\beta\lappeq3$. Also shown in
Fig.~\ref{fig:f4}
as a dashed line is the fine-tuning price that was imposed by the first round
of direct searches at LEP~1, which we model crudely by the
requirement that all charged and strongly-interacting sparticles
weigh $\gappeq45$ GeV.
Since these
early were much less constraining, they corresponded to 
a much smaller fine-tuning price, and 
we see that $\Delta \lappeq 30$ was
possible for all the values of $\tan\beta$ above the infra-red
fixed point. The dotted
line in 
Fig.~\ref{fig:f4} shows the fine-tuning price that may need to be paid if 
LEP~2 does not find any sparticles or a MSSM Higgs boson in future runs at 
centre-of-mass energies $\lappeq200$~GeV. We see that $\Delta$ could be 
increased significantly at low $\tan\beta$, principally as a result of the 
increase in the 
LEP~2 reach in $M_h$ to about 100~GeV. LEP~2 has already raised significantly 
the price of fine tuning, particularly at low $\tan\beta$, and the price for 
$\tan\beta \lappeq 2$ could become exorbitant if no discovery is made with the
remaining LEP~2 energy upgrades. If one assumes that the principal constraint 
imposed by the FNAL Tevatron Run~II will be $M_{1/2} \gappeq 150$~GeV and 
that, for example, the reach in $M_h$ will not greatly exceed that of LEP~2, 
the fine-tuning price would not increase at small values of $\tan\beta$, but 
there could be a marginally increased price at intermediate $\tan\beta$.

We have not included in the above analysis the requirement that the relic
neutralino density fall in the range $0.1<\Omega_\chi h^2<0.3$. As already 
mentioned, this may occur for $0.2\lappeq m_0/M_{1/2}\lappeq1$, with larger 
values of $m_0/M_{1/2}$ corresponding to unacceptably large values of
$\Omega_\chi h^2$. Looking at panels (d) of Figs.~\ref{fig:f1},\ref{fig:f2} 
and \ref{fig:f3} we see that this restriction on $m_0/M_{1/2}$ increases 
the fine-tuning price noticeably only when $\tan\beta\sim10$, resulting in 
a small increase in the global minimum value of $\Delta$. A complete 
implementation of the relic-density constraint could only increase still 
further the fine-tuning price, but such an analysis is beyond the scope of 
this paper.

\begin{figure}
\psfig{figure=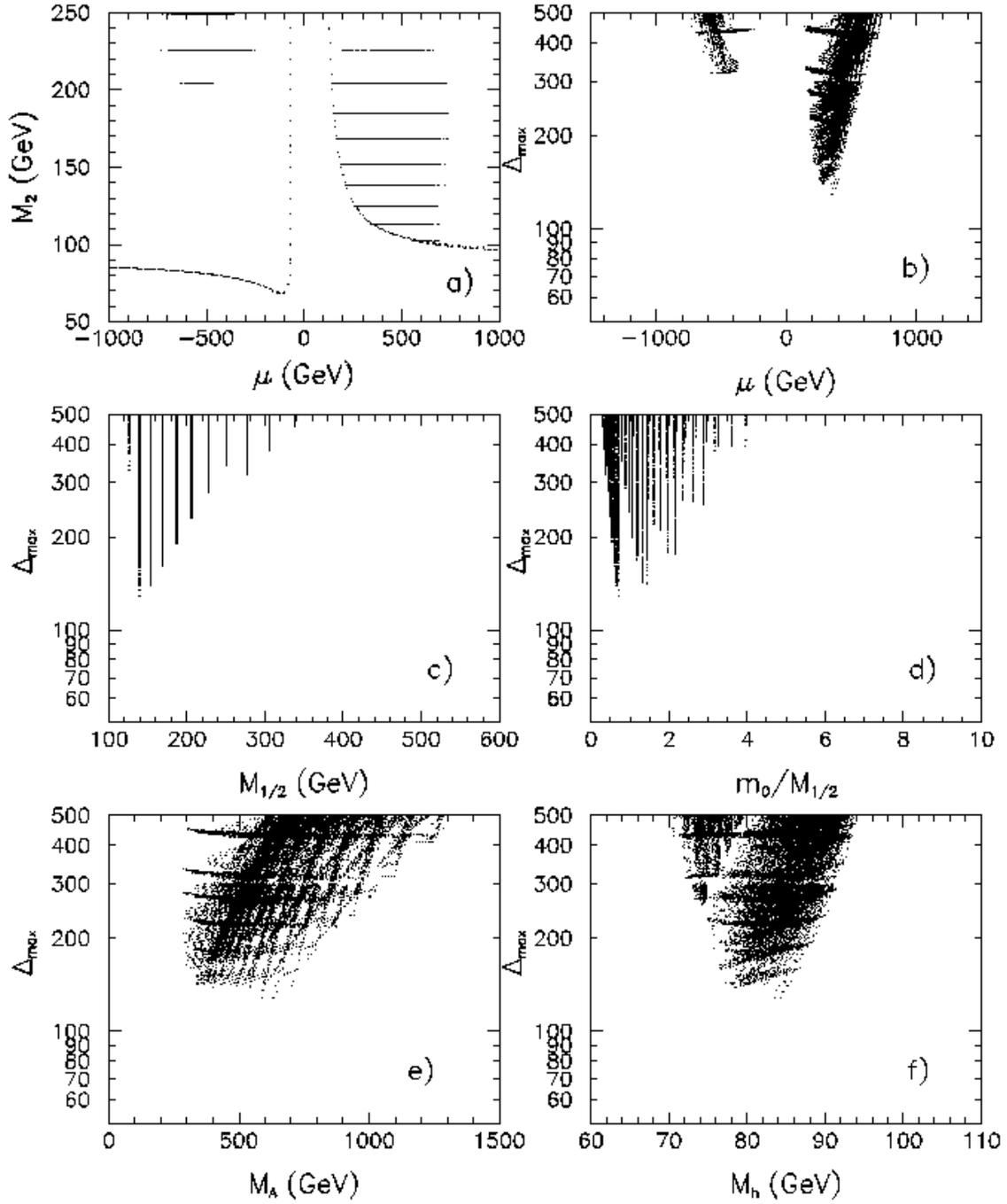,height=19.0cm}
\caption{As for Fig.~1, but now with non-universal input scalar masses for 
the Higgs multiplets: $m^2_{H_i} \neq m^2_0$, $i = 1,2$. In panel (a), only 
points with $\Delta_{max}<500$ are displayed.}
\label{fig:f5}
\end{figure}

We comment now on the possibility of non-universal input scalar masses for
the Higgs multiplets $m^2_{H_i}\neq m^2_0$, $i=1,2$. One would
think that the introduction of the two new parameters
$m^2_{H_i}$ must enable one to find parameter sets that require less fine
tuning. However, the appearance of the $m^2_{H_i}$ is accompanied by two
additional sensitivity parameters $a_i$
which must also be taken into account when evaluating $\Delta$.  We recall that
$\Delta$ is defined as the {\it maximum} of the sensitivities $\left\vert
(a_i / M^2_Z) (\partial M^2_Z / \partial a_i)\right\vert$ 
(\ref{eqn:e2}). This means that $\Delta$ could 
in principle even be {\it increased} by the introduction of 
the $m^2_{H_i}$. Fig.~\ref{fig:f5} shows our results for 
$\tan\beta=1.65$ with non-universal boundary conditions: after all cuts, 
the results are similar to those for universal scalar masses, though with a 
slight decrease in the minimal $\Delta$.
The most interesting point about non-universal Higgs boson mases is that
the region of small $M_{1/2}$ and small, negative $\mu$ is consistent with the 
requirement of the proper electroweak breaking but not with the experimental
cuts other than the limit on the chargino mass. After the cut $\Delta
\chi^2<4$ and/or $M_h>75$ this region is disallowed as for the universal
case. It would be allowed by all the cuts only after radical departure from 
the universality among the squark masses \cite{CACHOLPOWA},
which would increase the fine-tuning price.

Results for $\Delta$ for other values of tan$\beta$ and choices of
data sets are shown in the right panel of Fig.~\ref{fig:f4}.
Although there are differences in detail, the general trends are similar
to those for the universal case shown in the left panel. We conclude that
increasing the number of parameters in this way does not reduce
significantly the fine-tuning price.

We have stressed already that fine tuning is a subjective issue: there is no
unambiguous method for evaluating it, and there is no objective criterion 
for deciding when the price is too high. Moreover, if one or more of the 
parameters $a_i$ is fixed by some external condition such as some more 
sophisticated theoretical assumption, $\Delta$ may well be reduced. We can 
illustrate this point by calculating $\Delta$ under the hypothetical 
assumption that some theory predicts a relation between a pair of
the five parameters, so that we have now only four independent input 
parameters. For example, if there is a linear relation\footnote{The 
requirement of the electroweak symmetry breaking selects approximately such a 
subspace in the parameter space \cite{CAOLPOWA}.} between $\mu$ and $M_{1/2}$, 
we find results that are qualitatively similar to those shown in 
Fig.~\ref{fig:f1} for tan$\beta = 1.65$, but with the minimum value of 
$\Delta$ reduced by a factor $\sim 4$. We have also found that $\Delta$ could 
be reduced by postulating a linear relation between $\mu$ and $B_0$, but this 
is mainly for low values of $M_h$ that are apparently excluded by the latest 
LEP~2 limits~\cite{LEPC183}.

In this paper we have made a first attempt to pose the experimental 
constraints on fine tuning in an objective way. We have seen that LEP~2 has 
raised the fine-tuning price by a significant factor, particularly at low 
$\tan\beta$ close to the infra-red fixed point. We have seen that important 
r\^oles in this price rise has been played by precision measurements,
the observation of $b \rightarrow X_s \gamma$ decay and to  
some degree the non-observation of the lightest MSSM Higgs boson. The price 
could rise again if the Higgs boson is not discovered with subsequent runs of 
LEP~2 at higher energies. Moreover, the price is not reduced by postulating 
non-universal boundary conditions for the Higgs scalar masses, and could be 
further increased if one imposes an astrophysical requirement on the relic 
neutralino density.

Personally, we do not find the present fine-tuning price too high, particularly
for $\tan\beta \gappeq 2.5$.  The price rise at low $\tan\beta$ does diminish
somewhat the attraction of the infra-red fixed point.  However, this is a
luxury model with added features, so the reader may be prepared to pay a 
higher price for it!  Alternatively, some more predictive theory may
correlate some of the five MSSM parameters that are currently regarded
as independent, which may reduce $\Delta$ significantly.

\vskip 0.2cm
{\bf Acknowledgments} ~~
P.H. Ch. would like to thank the CERN Theory Division for hospitality
during the completion of this work. The work of P.H. Ch. and of S. P.
was partly supported by Polish State Commitee for Scientific Research under
grant 2 P03B 040 12 (for 1997-98).

\end{document}